\documentclass[9pt]{article}
\usepackage{latexsym}
\usepackage{graphicx}
\usepackage{amsfonts,amssymb}

\begin{document}

\title{\textbf\texttt{{The attack tolerance of community structure in complex networks}}}
\author{Jie Cheng$^{1}$, Xiaojia Li$^{1}$, Zengru Di$^{1}$, Ying Fan$^{1}$\footnote{Author for correspondence: yfan@bnu.edu.cn}
\\ \emph{1. Department of Systems Science, School of Management,}\\
    \emph{Beijing Normal University, Beijing 100875, P.R.China}}

\maketitle

\begin{abstract}
Robustness is an important property of complex networks. Up to now,
there are plentiful researches focusing on the network's robustness
containing error and attack tolerance of network's connectivity and
the shortest path. In this paper, the error and attack tolerance of
network's community structure are studies through randomly and
purposely disturbing interaction of networks. Two purposely
perturbation methods are designed, that one methods is based on
cluster coefficient and the other is attacking triangle.
Dissimilarity function $D$ is used to quantify the changes of
community structure and modularity $Q$ is used to quantify the
significance of community structure. The numerical results show that
after perturbation, network's community structure is damaged to be
more unclear. It is also discovered that purposely attacking damages
more to the community structure than randomly attacking.
\end{abstract}



\section{Introduction}
In recent years, more and more systems in many fields are depicted
as complex networks. Robustness, as an important function of many
systems, is studies under framework of complex networks recently.
The past researches mainly focus on the the topological aspects of
robustness and are usually done either by removal of nodes, random
failure of nodes or targeted attack, or by replacing a sector of
edges or adding new edges\cite{Reka Albert}. Network with special
structure, for instance scale-free network, is found to be less
robust to targeted attack and more robust to random failure. The
robustness is usually measured by either the size of the largest
connected component or the length of the shortest path between pairs
of nodes. Some of the investigations try to find effective way to
improve robustness of real networks against attack.

During the studies on networks, lots of evidences show that there
are communities in social networks, metabolic networks, economic
networks\cite{biological,InterbankMarket,metabolic,Nature,Bioinformatics,jazz}
and so on. Community structure is an important character to
understand the functional properties of complex networks. For
instance, in social networks, communities can formed depending on
careers or ages. In food web, communities may reveal the subsystem
of ecosystem \cite{Williams}. In biochemical or neural networks,
communities may correspond to functional groups
\cite{Bioinformatics}. In the world wide web, the community analysis
have found thematic groups \cite{Eckmann,functional}. Email network
can be divided into departmental groups whose work is distinct and
the communities reveal organization structures \cite{emaila,emailb}.
Moreover, the study of dynamics in complex networks shows that
vertices belonged to the same community reach synchronization easily
\cite{reveals}. Thus deep understanding on community structures will
make us comprehend and analyze the characteristics of systems
better. Most research about the community structure mainly focus on
the algorithm of detecting the community structure in networks
\cite{biological,Radicchi}.

Robustness of community structure hasn't been paid enough attention.
Recently, Brian Karrer et al. \cite{robustness}proposed a random
perturbation method with rewiring edges randomly, and studies the
robustness of community structure, which is an error tolerance
study. In \cite{robustness}, it is proposed that robustness study
can be another method of measuring the significance of community
structure. The attack tolerance of community structure hasn't been
studied yet, which is mainly focused in this paper. Based on the
idea in \cite{robustness}, we propose two targeted methods and
investigated their effect on community structure in networks. This
paper is organized as following. In section \ref{Attack Methods on
network}, we introduce three network perturbation ways, including
one random disturbing way. In section \ref{Robustness of Community
Structure}, random and attack tolerance of community structure are
investigated through dissimilarity functions $D$. We disturb
networks using targeted attack and random attack method on
artificial and real network and numerical results are analyzed.
Section \ref{Modularity and Edge-clustering coefficient}, attack's
effect on the significance of community structure and network's
topology character is explored through modularity $Q$'s and average
cluster coefficient $C$. It is found that targeted attack cause
serious damage to network's structure. The attack is simulated both
on artificial and real network and numerical results are analyzed.
Section \ref{conclusion} is a conclusion.

\section{Attack Methods on network}\label{Attack Methods on network}
In network, triangles as an basic unit plays important role in
communities structure.  There are divisive algorithms basing on the
triangle structure to find community structure in networks, such as
the algorithm invited by Filippo Radicchi, et al. \cite{coefficient}
and good results are obtained. It is found that edges connecting
nodes in different communities are included in few or no triangles.
Edge-clustering coefficient is introduced first \cite{coefficient}
for the edge-connecting node $i$ to node $j$:
\begin{equation}
C_{ij}^{3} = \frac{z_{ij}^{3}}{min[(k_{i}-1),(k_{j}-1)]}
\end{equation}

where $z_{ij}^{3}$ is the number of triangles built on that edge and
$min[(k_{i}-1),(k_{j}-1)]$ is the minimal possible number of them.
Edges connecting nodes within communities are included in more
triangles and tend to have large values of $C_{ij}^{3}$. Our first
targeted attack method is attacking edges with the largest
$C_{ij}^{3}$, and second targeted attack method is attacking edges
which is included in the largest triangles. Initially a network with
$N$ nodes and $M$ edges is given.

The targeted attack method based on edge-clustering coefficient( EC
method in short) is as followings.
\begin{enumerate}
\item Compute $C_{ij}^{3}$ on every edge and ascending the
edge according to the cluster coefficient.
\item  Remove the former $\alpha$*$M$ edges from the network.
\item Go through every pair nodes ($i$,$j$) and rewire $\alpha$*$M$ edges
randomly selected from the network.
\end{enumerate}

It can be known that no edge is moved as $\alpha$ = 0. As $\alpha$ =
1, all edges are moved. The number of nodes and edges keeps constant
ever if moved.

The targeted attack method based on triangles ( T method in short),
is as followings.
\begin{enumerate}
\item Compute $z_{ij}^{3}$on every edge and  ascending the
edge according to $z_{ij}$.
\item calculate the total number of $t$ = $z_{ij}^{3}\neq0$.
and remove the former $\alpha_{1} \ast t$ edges from the network.
\item Randomly add $\alpha_{1} \ast t$ edges to the network.
\end{enumerate}
In T method, $t$ represents the total number of edges included in
triangles. It can be known that no edge is moved as $\alpha_{1}$ =
0. As $\alpha_{1}$ = 1, $t$ edges are moved. For comparison to other
perturbation method, here we still use $\alpha$ to represent the
ratio of number of edges actually moved to the total number of
edges. $\alpha = \frac{\alpha_{1} \ast t}{M}$. The number of nodes
and edges keeps constant ever if moved.

The random attack methods ( R method in short) is designed here for
comparison to the targeted attack.
\begin{enumerate}
\item remove $\alpha \ast M$ edges from the network randomly.
\item Randomly add $\alpha \ast M$ edges to the network.
\end{enumerate}.
The main difference between above two targeted attack methods is
that the second methods disturbing the only edges contained in
triangles purposely rather than all of the edges in the network.
Thus the radices of the perturbation are changed. In the second
methods if $\alpha = 0$, no edges are rewired. However, if
$\alpha=1$, all of the edges that consist of triangles are rewired,
rather than all of the edges in the network. During the perturbation
the average degree and size $N$ of the network keep constant. The
attack methods which is different from the method in
\cite{robustness}, is that our perturbation scheme generates
networks in which the expected degrees of vertices aren't the same
as the original degrees.

\section{Robustness of Community Structure}\label{Robustness of Community Structure}
\subsection{Quantification of variation of Community Structure}
Quantifying difference of community structures is not a new
problem\cite{Danon,Kuncheva,Fred}. The community structure divided
by algorithm is usually compared with the actual communities if
known. When analyzing the precision of a certain algorithm, the
comparison of different community structures is also needed. Thus
several methods for measuring similarities or differences between
divisions of community structure have been designed. Function $D$
\cite{dissimilarity}is a simple and efficient measurement to the
difference between two community division and is chosen here as an
measurement of robustness of community structure.

In this section, based on the three perturbation methods proposed
above, we measure the variation of robustness of community structure
after perturbation using dissimilarity function $D$. While in
Karrer's study, he analyzed the robustness using function $V$ based
information entropy \cite{robustness}. The reason we choose $D$ is
that $D$ reflects the same character of variation of information of
community structure as $V$ does, and  $D$ can be normalized to 1.
When doing the computer experiments, it is found that $D$ is more
sensitive than $V$.

The idea of dissimilarity function $D$ is introduced by
\cite{dissimilarity}. Discuss the similarity and dissimilarity of
two sets $A$ and $B$ that defined as the subsets of $\Omega$.
Similarity is expressed by $A\cap B$, and dissimilarity corresponds
to be $\left(A\cap\bar{B}\right)\cup\left(\bar{A}\cap B\right)$. The
normalized similarity and dissimilarity can be represented as
\begin{equation}
\left\{
\begin{array}{l}
s = \frac{\left|A\cap B\right|}{\left|A\cup B\right|}\\
\\
d = \frac{\left|\left(A\cap\bar{B}\right)\cup\left(\bar{A}\cap
B\right)\right|}{\left|A\cup B\right|}
\end{array}\right.
\end{equation}

Discussing two particular divisions of a network, each of them have
many communities, and we assuming that both of them have $k$
communities. First, construct the correspondence between communities
that from different sets, which makes them have biggest similarity.
Second, calculate the dissimilarity of each pair of the subsets. And
then using the dissimilarities of all subsets to calculate the
integer sets' dissimilarity.

\begin{equation}
D=\frac{\sum_{i=1}^{k}{d_{X_{i}Y_{i}}}}{k}
\end{equation}

However, in most cases, two community structures do not have the
same number of communities, which means not every subset has
correspondence subset. To solve this problem,the subset $X_{i}$ that
has no correspondence, correspond with $\Phi$. The $k$ equal to the
larger number of communities. Under this definition, the maximum
value of $D$ is 1 and minimum value of $D$ is 0, where
$\left(0,1\right)$ means no and largest differences respectively.

\subsection{Numerical results of Robustness of Community Structure}
\begin{figure}[th]
\centering\includegraphics[width=4cm]{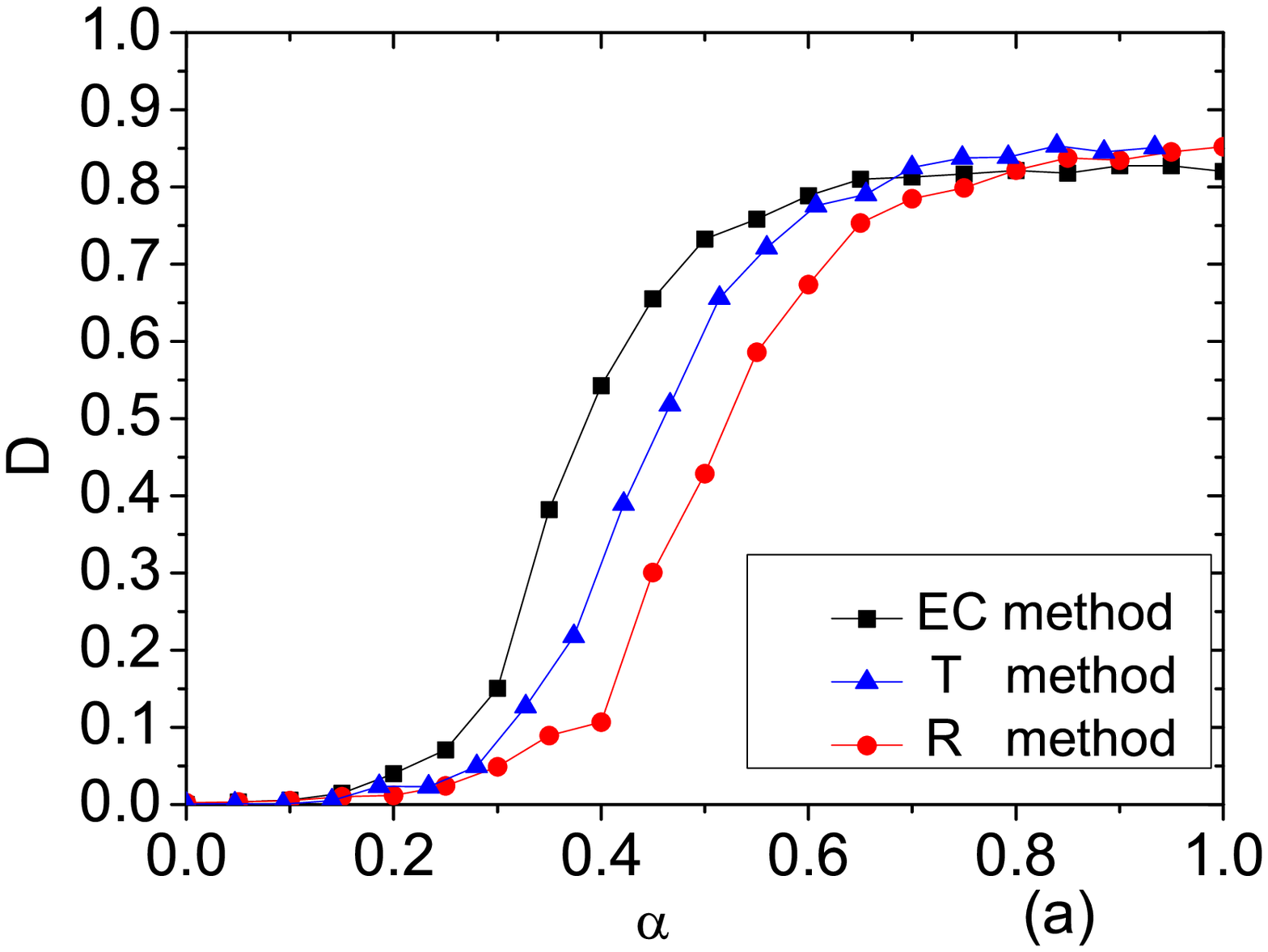}
\includegraphics[width=4cm]{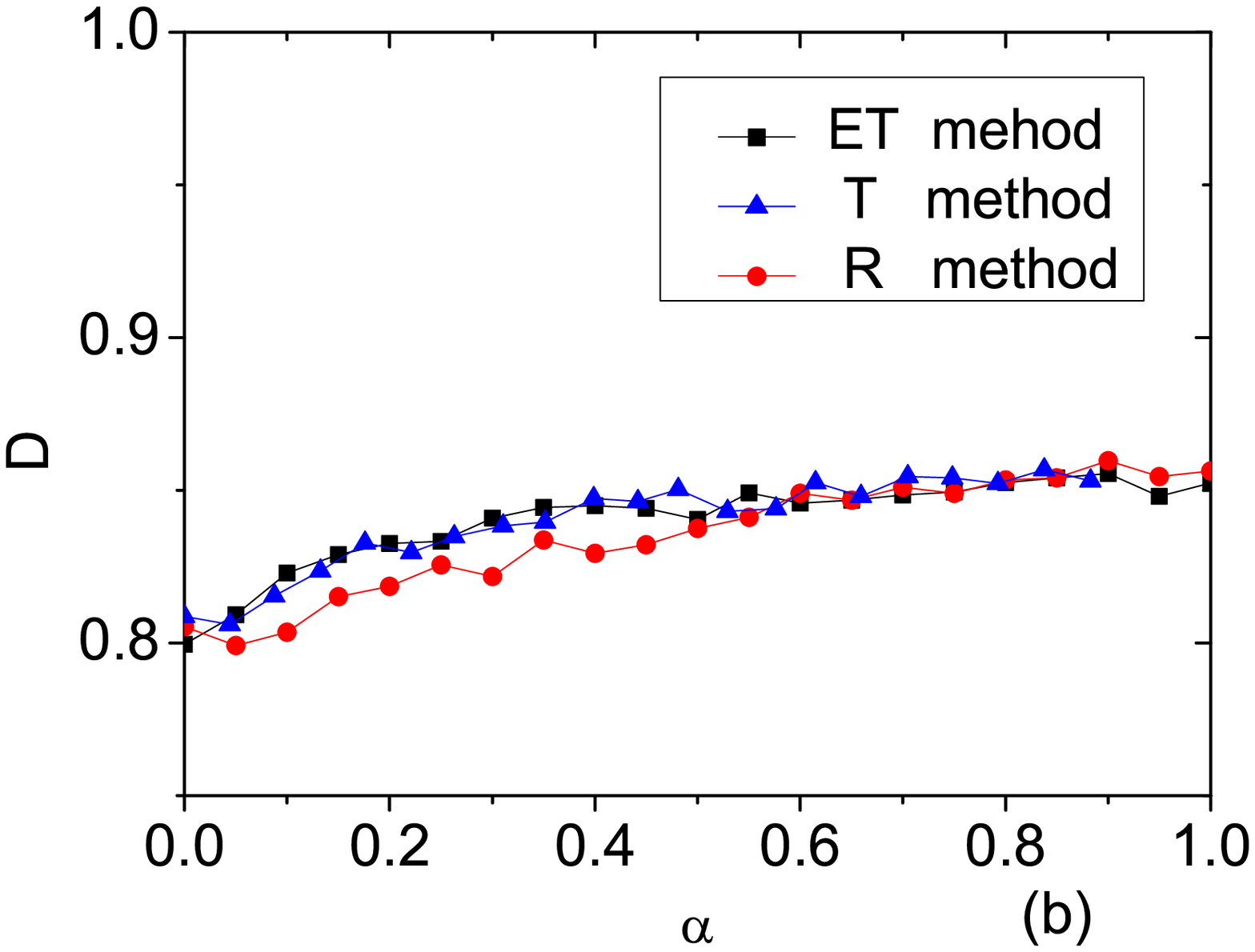}
\includegraphics[width=4cm]{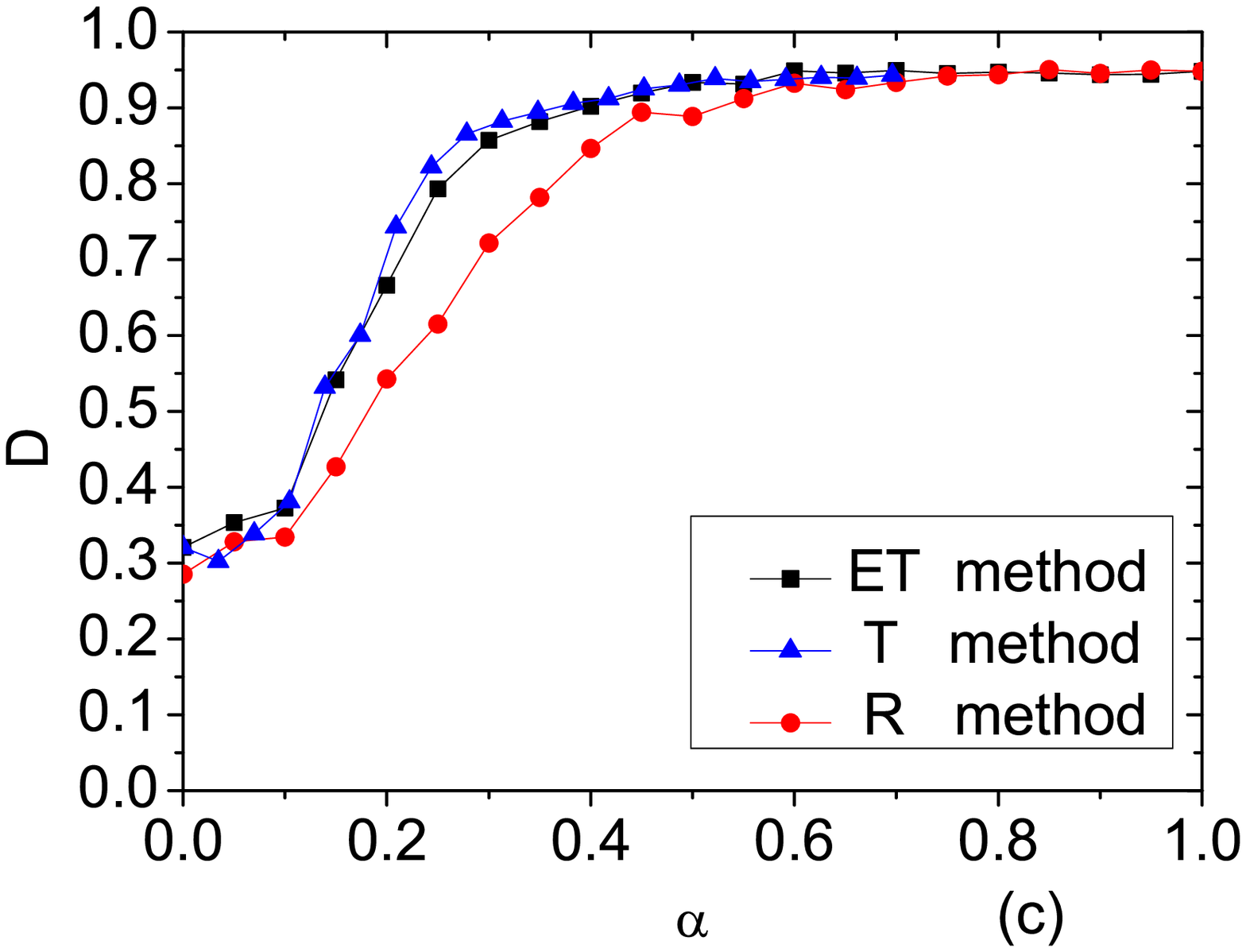}
\caption{Value of Function $D$ versus $ \alpha$. Three attack
methods are simulated on the arbitrary network benchmark.
(a)$k_{out}$ = 2 on GN benchmark;(b)$k_{out}$ = 10 on GN
benchmark;(c) LFR benchmark, $N$ = 500, $\overline{k}$=10, $p(k)\sim
k^{-\gamma}$, $\gamma = 2.5$, $p(s)\sim k^{-\lambda}$, $\lambda$ =
2.0, mixing parameter $\mu$ = 0.3.}\label{fig1}
\end{figure}

\begin{figure}[th]
\centering\includegraphics[width=4cm]{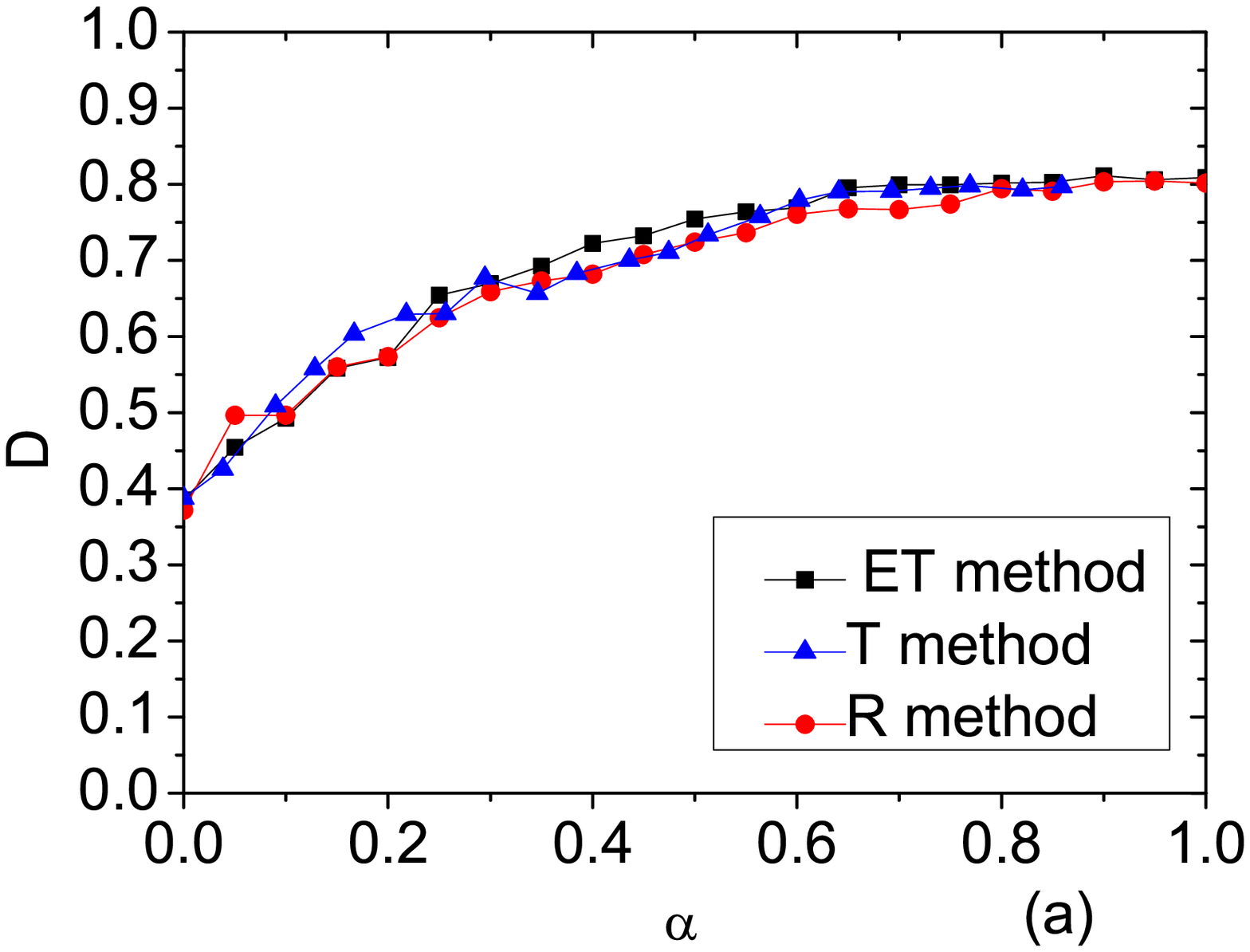}
\includegraphics[width=4cm]{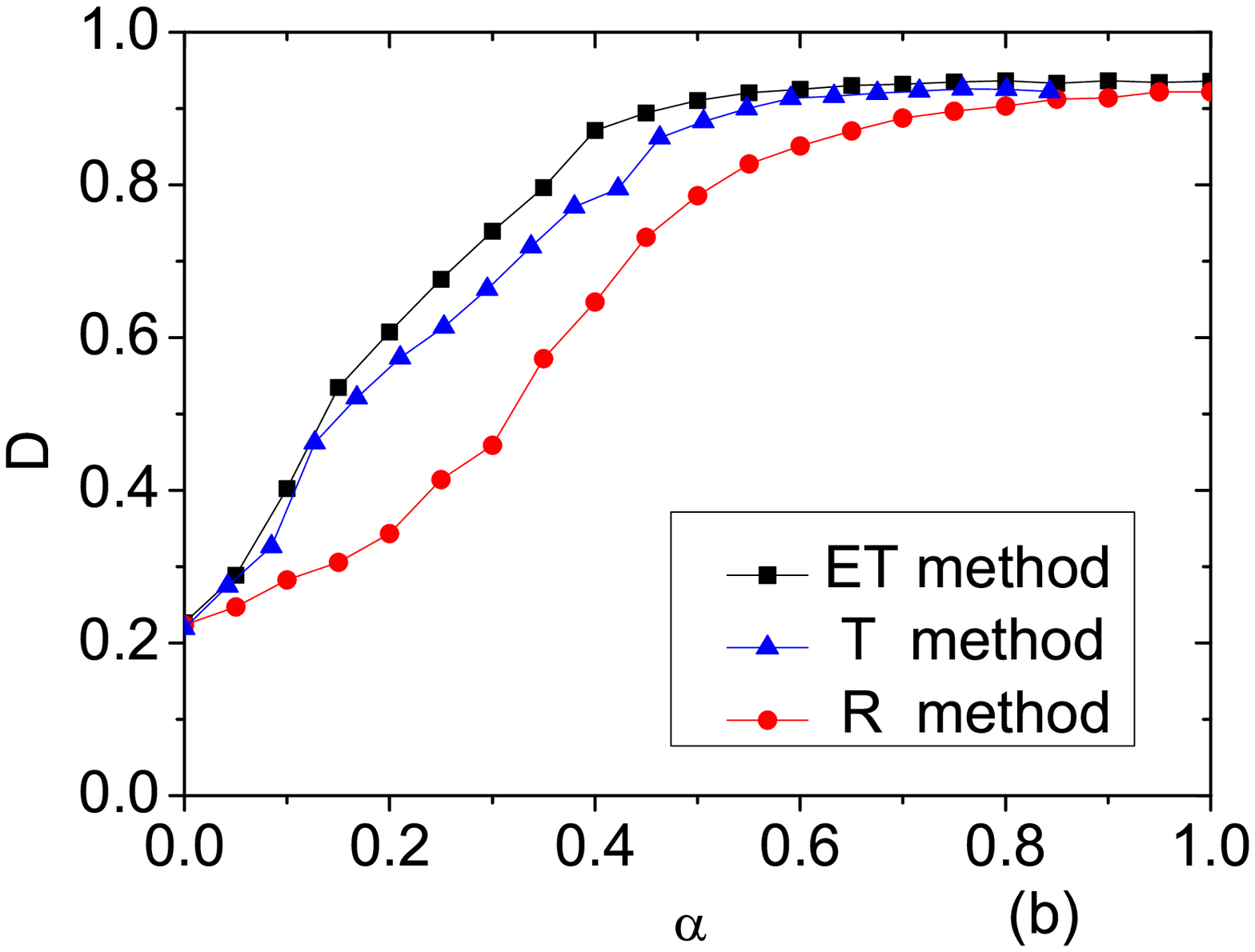}
\includegraphics[width=4cm]{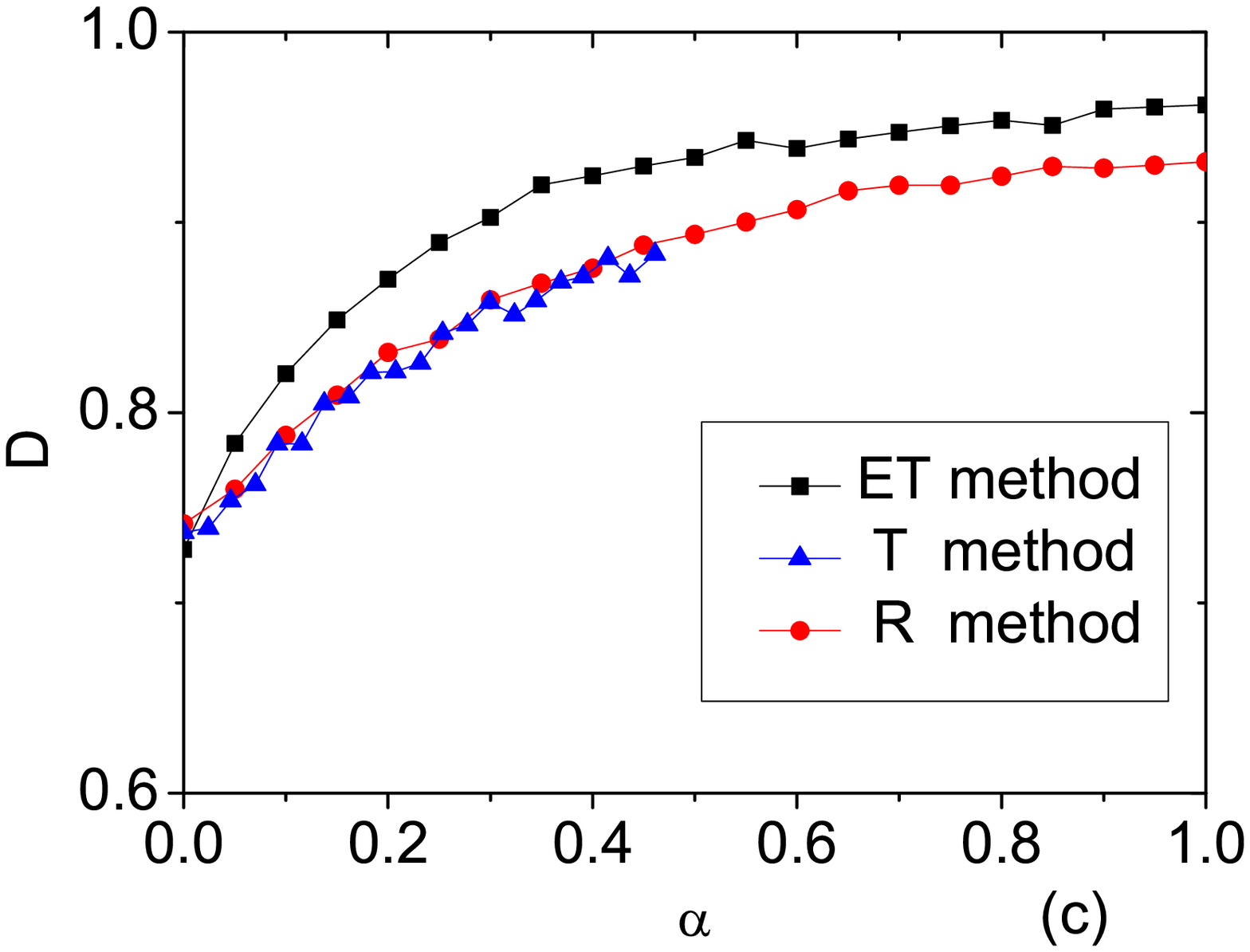}
\caption{ Value of Function $D$ versus $\alpha$. Three attack
methods are simulated on real network .(a) Three attack methods are
simulated on Karate club network. (b) Three attack methods are
simulated on  football network. (c)Three attack methods are
simulated on econo-physicist network.}\label{fig2}
\end{figure}

For a given network, now we have all the components to analyze the
robustness. First, we get the community structure $C$ of this
network by any algorithm existing now. Here we use EO algorithm to
detect the community structure for its good character \cite{EO}.
Then, we disturb the original network with the methods proposed in
above section separately, and get the new community structure $C'$.
And then, measure the varieties of community structure by function
$D$ which have introduced in section \ref{Attack Methods on
network}. Since the disturbing methods including some choosing by
random, we will repeat the second step for some times, and get the
average value of variation to make sure it not be impacted by
special cases. And the whole process is done many times
concomitantly with the change of parameter $\alpha$ from 0 to 1.

To test the efficiency of targeted attack methods proposed here,
three attack methods are simulated respectively on artificial
networks generated by GN benchmark and LFR benchmark, and real
network including Karate network, football network, and
econophysicist collaboration network.

As first, we apply the methods on benchmarks of Girvan-Newman (GN)
proposed in \cite{128}. In GN benchmark, homogeneous networks are
generated  and used widely in the evaluation of community detection
algorithms. These networks consist of 128 vertices divided into 4
communities of 32 nodes each. Every node is connected on average
with $\langle k_{in} \rangle$ nodes of its own group and $\langle
k_{out} \rangle$ of the rest of the network. The total degree of
each node is equal to $k = k_{in}+ k_{out}$ and always kept constant
to 16. As the average number $k_{out}$ of between-group connections
per vertex is increased from zero, the community structure in the
network, stark at first, becomes gradually obscured until, at the
point where between- and within-group edges are equally likely, the
network becomes a standard Poisson random graph with no community
structure at all. Here, two cases that $k_{out}$= 2 and $k_{out}$ =
10 are used.

Fig.\ref{fig1}(a,b) shows the results of the application of our
analysis method to graphs of this type. The figure shows the value
of the variation of community structure $D$ as a function of the
parameter $\alpha$ that measures the amount of perturbation. For
small value of $k_{out}$ that $k_{out}$ = 2 the variation of
community structure increases faster under two targeted attacks than
under random attack as a function of $\alpha$  as shown in
Fig.\ref{fig1}(a). As we can see, the variation of community
structure $D$ starts at zero when $\alpha$ =0, as we would expect
for an unperturbed network, rises rapidly, then levels off as
$\alpha$ approaches its maximum value of 1. The curves of targeted
methods depart significantly from that of random method, indicating
that the community structure discovered by the algorithm is less
robust against targeted perturbation. Furthermore, the curve of ET
method depart significantly from that of T method, indicating that
the community structure discovered by the algorithm is relatively
fragile against the ET method.

We can find that the curves represent the three methods don't have
the same length. The reason is the total edges that are changed in
different methods are not the same. For R method ET method, all of
the edges in the network can be moved. For T method disturbing
towards triangles, the edges that can be disturbed are the edges
that forming triangles, number of which may be much fewer than the
total number of the edges.

Large values of $k_{out}$ in GN benchmark that generates network
with obscure community structure, and $k_{out}$ = 10 here.  As shown
in Fig.\ref{fig1}(b), $D$ keeps a high value and almost unchanged as
$\alpha$ changes from zero to one under three attack methods,
indicating that the attack tolerance and error tolerance of obscure
community structure is closed to each other. It is necessary to
point out that $D$'s minimal value is not always exact zero when
$\alpha$ = 0, which is determined by the EO algorithm.

Then, on benchmarks of Lancichinetti et al. (LFR) in
\cite{Lancichinetti}, the methods is applied to disturb the network
generated. LFR is a generalization of the GN benchmark to
heterogeneous group sizes and graph degree distribution. Groups are
also a priori fixed with the degrees and the community sizes
following a power-like distribution. As before, nodes have $k_{in}$
connections within its own group and $k_{out}$ edges linking
elsewhere. For investigation of robustness of community structure,
networks with significant community structure is needed here, and
the parameters of LFR benchmark is set as following. The average
degree $\langle k \rangle$ = 10, and size is 500. The degree
distribution follows a power-law $P(k) \sim k^{\gamma}$, with
$\gamma$= 2.5; and the community sizes distribution follows a
power-law $P(k) \sim k^{\beta}$, with $\beta$= 2. The mixing
parameter $\mu= k_{out}/k$ indicates the "strength" of the
communities, and is set to be 0.3 here. The maximal community size
is 80, and the minimal is 20. Under such parameters, the network
with clear community structure can be generated.

For the LFR benchmark, as shown in Fig.\ref{fig1}(c), the curves of
targeted methods also depart significantly from that of random
method, indicating that the community structure discovered by the
algorithm is fragile against targeted perturbation. With moving a
small amount of edges with high $C$ or high number of triangles, the
community structure changes more than with random moving edges. The
results on the three artificial networks suggest that targeted
methods is more efficiently destroying the community structure than
random attack.

Turning now to real-world networks, we have tested our method on
examples mainly including social networks. A selection of results
are shown in Fig.\ref{fig2}.

Fig.\ref{fig2}(a)  shows the curve of variation of community
structure as a function of $\alpha$ for one of the best studied
examples of community structure in a social network, the ¡°karate
club¡± network of Zachary  \cite{karate}. The vertices in this
network represent members of a karate club at a US university in the
1970s and the edges represent friendship between members based on
independent observations by the experimenter. The network is widely
believed to show strong community structure and repeated studies
have upheld this view.

The black(square) and blue(triangle) points in the figure show the
variation of community structure under targeted attacks while the
red points show the results under random attack. It is clear in this
case that the community structure is essentially the same robust
against random perturbation with targeted perturbation.

Then we apply the methods to the football network \cite{Girvan} are
shown in Fig.\ref{fig2}(b). In the network of American college
football teams. is a representation of the schedule of Division I
games for the 2000 season: vertices in the graph represent teams
(identified by their college names) and edges represent
regular-season games between the two teams they connect. The network
contains 115 nodes, 613 edges and is proved to have significant
community structure. The curves of targeted methods also depart
significantly from that of random method, indicating that the
community structure discovered by the algorithm is relatively less
robust against targeted perturbation. And for a certain possibility
that edges have been disturbed, the variation of community structure
caused by the methods we introduced are larger than variation of
community structure caused by the random disturbing.

The result of robustness of econophysicists collaboration network
\cite{econophysicist} are shown in Fig.\ref{fig2}(c). In the
econophysicists collaboration network, notes represent
econophysicists, the edges represent their collaboration
relationship. And we analyze the largest component of it, which
contains 271 nodes. For the same reason that different methods
disturbing different edges, the curves in Fig.\ref{fig2}(c) are not
in the same length. The curves of ET method departs significantly
from that of R method, indicating that the community structure
discovered by the algorithm is relatively fragile robust against ET
perturbation.

Comparing the analysis in the three real networks, we can find that
the disarrange methods we proposed are always more efficiency than
random disturbing. The results shows that under the same
perturbation strength, targeted attack methods is more efficient
than random attack method. Targeted methods always cause more damage
to the original community structure. These results indicate that
when investigating robustness of community structure, targeted
attacks based on edge-clustering coefficient or triangles can be
more efficient methods.

\section{Modularity and Edge-clustering coefficient}\label{Modularity and Edge-clustering coefficient}
In above sections, the robustness of community structure is
discussed through targeted and perturbation on network, and it is
found that the community structure is fragile to targeted attack. In
this paper, we mainly focus on the perturbation's effect on the
topology character of network, such as modularity function and
edge-clustering coefficient, which are related to community
structure.
\subsection{Modularity function}\label{funcion}
Modularity function $Q$
\cite{newman greedy better} now is implemented widely to measure the
significance of network's community structure.

\begin{equation}
Q=\frac{1}{2m}\sum_{ij}[A_{ij}-\frac{k_ik_j}{2m}]\delta(c_i,c_j)
\end{equation}
where $A_{ij}$ is an element of the adjacency matrix of the network,
$A_{ij}=1$ if there is an edge between node $i$ and node $j$,
otherwise $A_{ij}=0$. $k_i$ represents the degree of the vertex $i$,
which is defined to be the number of edges connected to node $i$,
$m=\frac{1}{2}\sum_{ij}A_{ij}$ is the number of edges in the whole
graph, and $c_i$ shows that vertex $i$ belongs to community $c_i$.
If the community structure is divided properly, the faction of edges
within communities should be large than the expect for the
randomized network. The larger value of the function $Q$ means the
better community structure. For a given process people can calculate
$Q$ for each split of a network into communities, and there are only
one or two local peaks. The position of these peaks usually
correspond closely to the expected divisions.

\begin{figure}[th]
\centering\includegraphics[width=4cm]{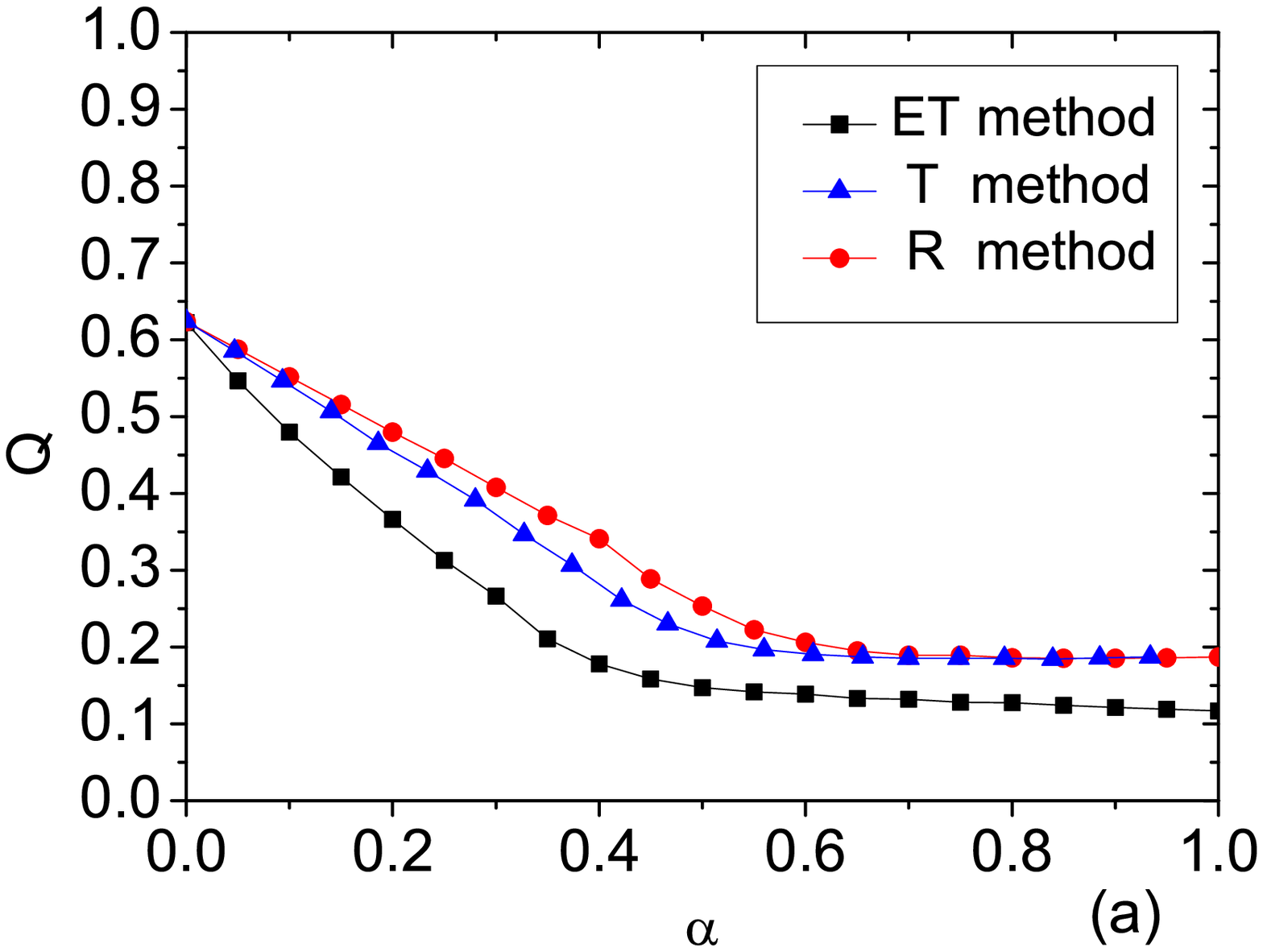}
\includegraphics[width=4cm]{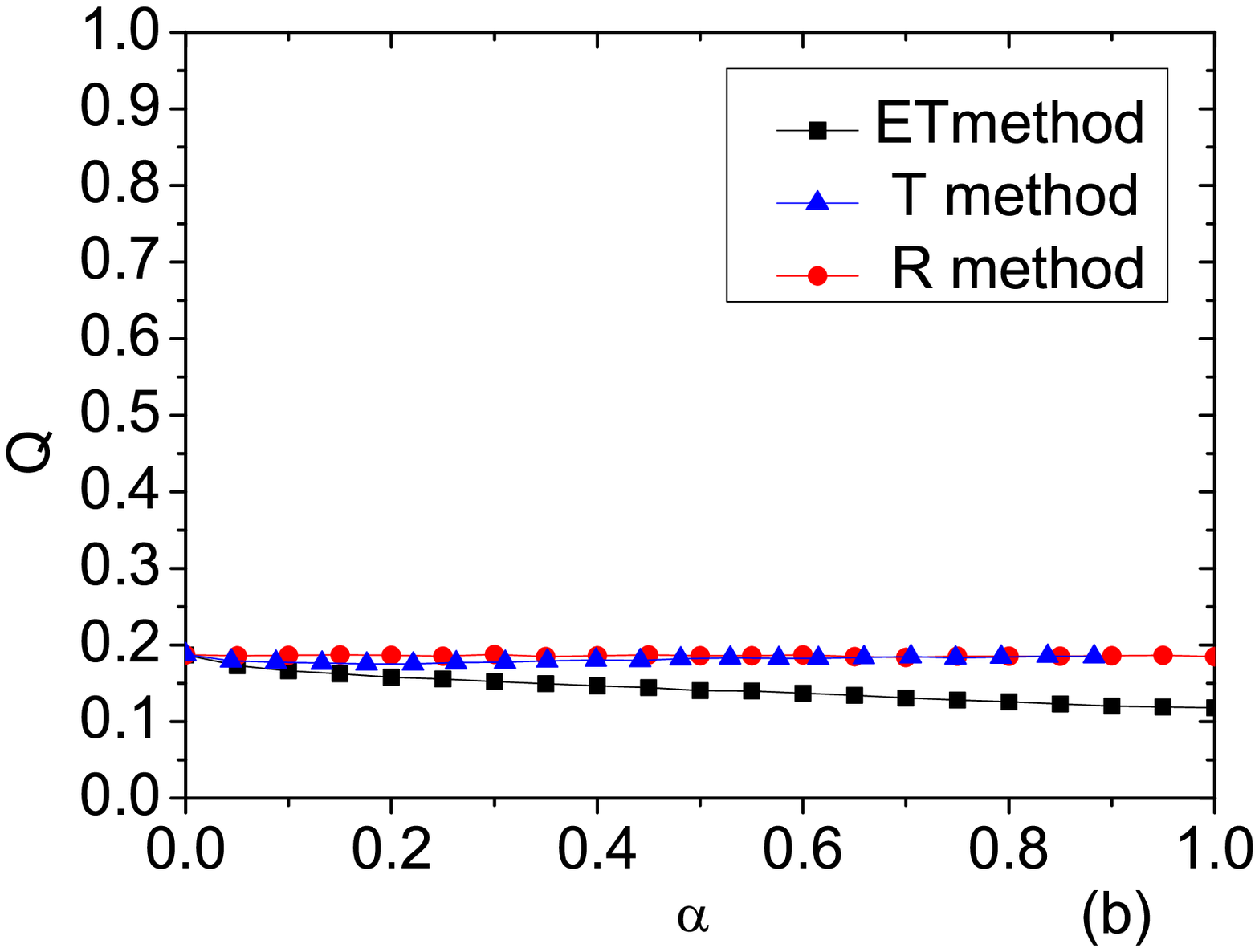}
\includegraphics[width=4cm]{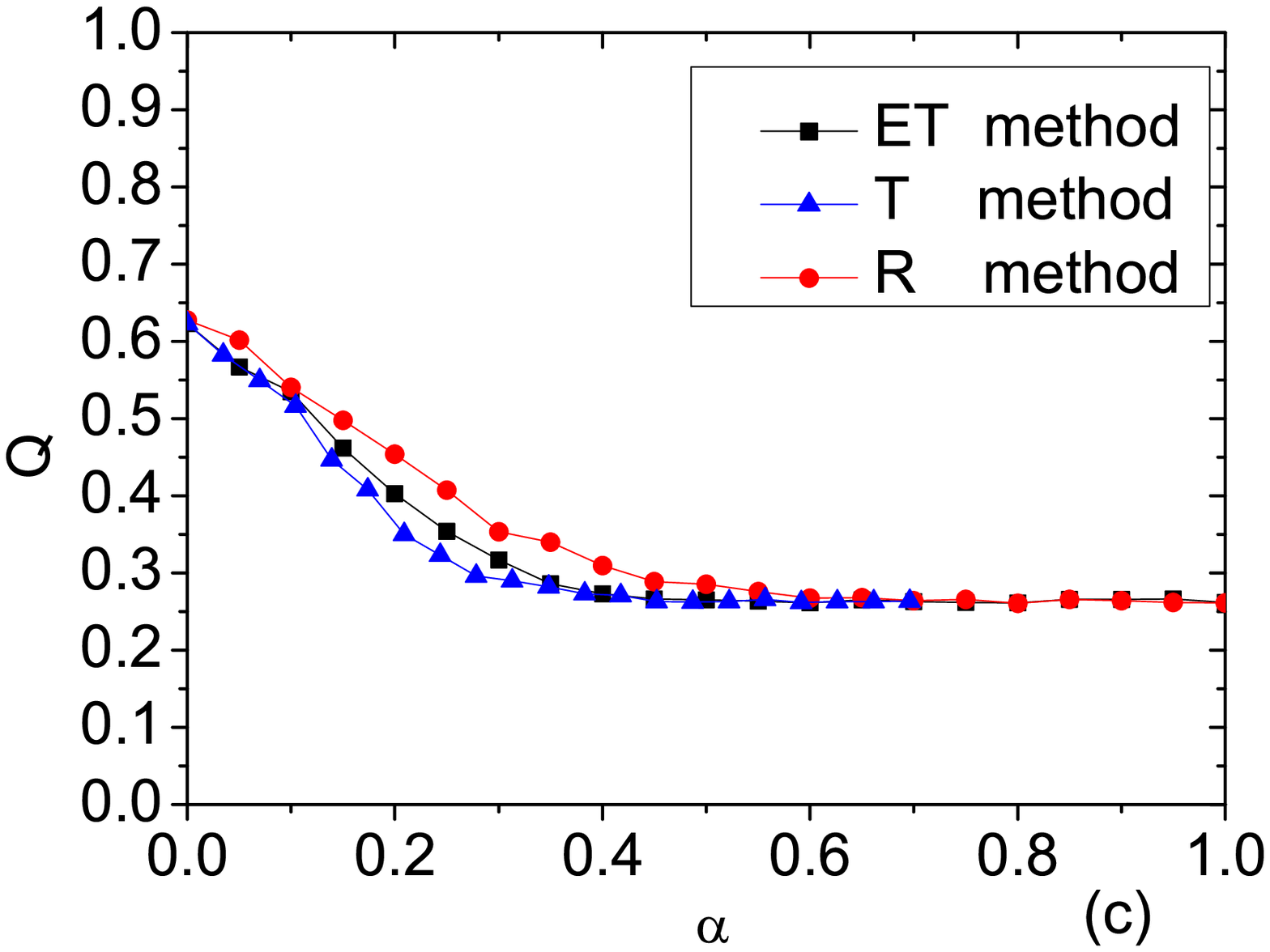}
\caption{ Value of Modularity $Q$ versus $\alpha$. Three attack
methods are simulated on the arbitrary network
benchmark.(a)$k_{out}$ = 2 on GN benchmark;(b)$k_{out}$ = 10 on GN
benchmark;(c) LFR benchmark, $N$ = 500, $\overline{k}$=10, $p(k)\sim
k^{-\gamma}$, $\gamma = 2.5$, $p(s)\sim k^{-\lambda}$, $\lambda$ =
2.0, mixing parameter $\mu$ = 0.3. }\label{fig3}
\end{figure}

\begin{figure}[th]
\centering\includegraphics[width=4cm]{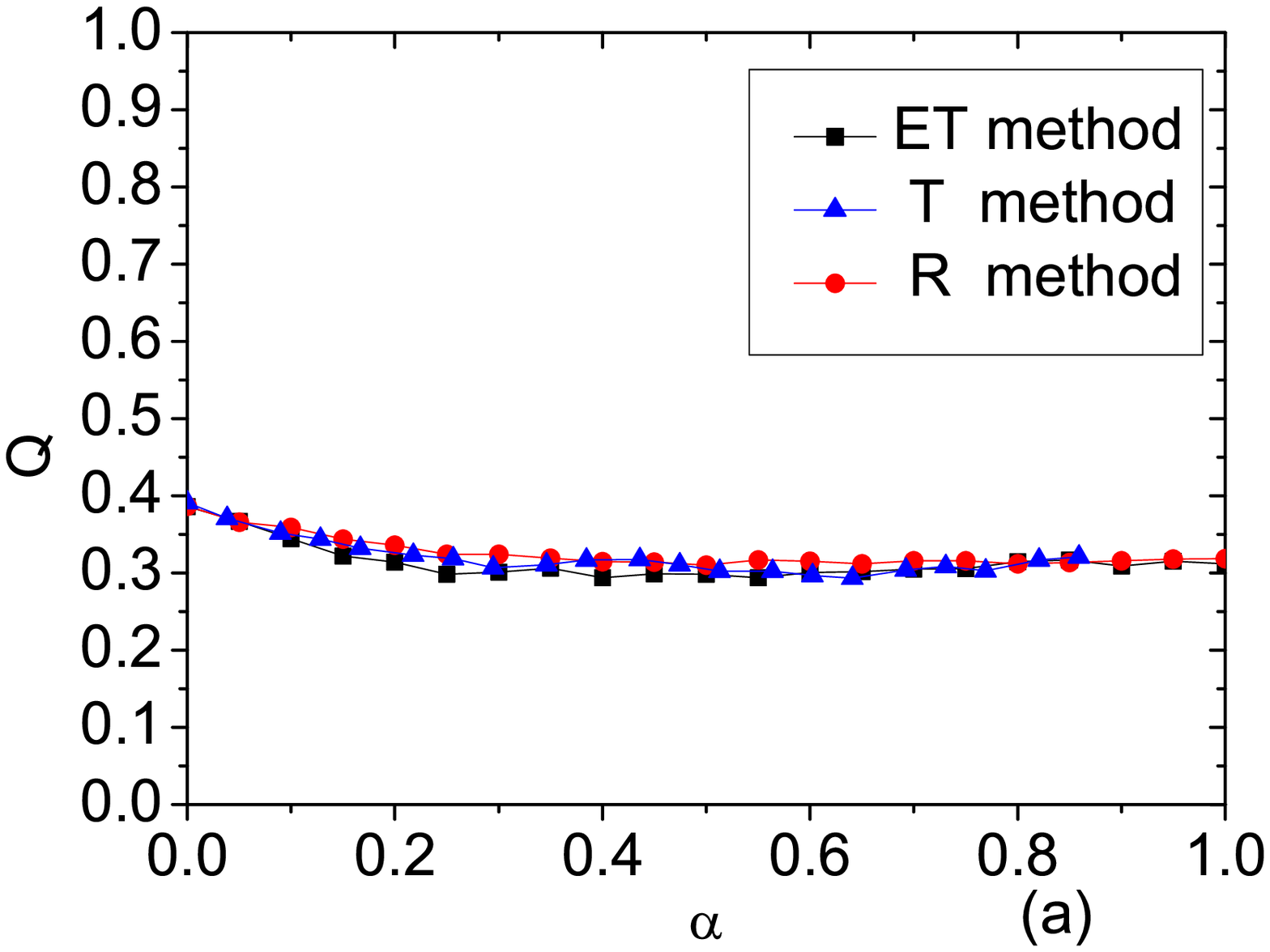}
\includegraphics[width=4cm]{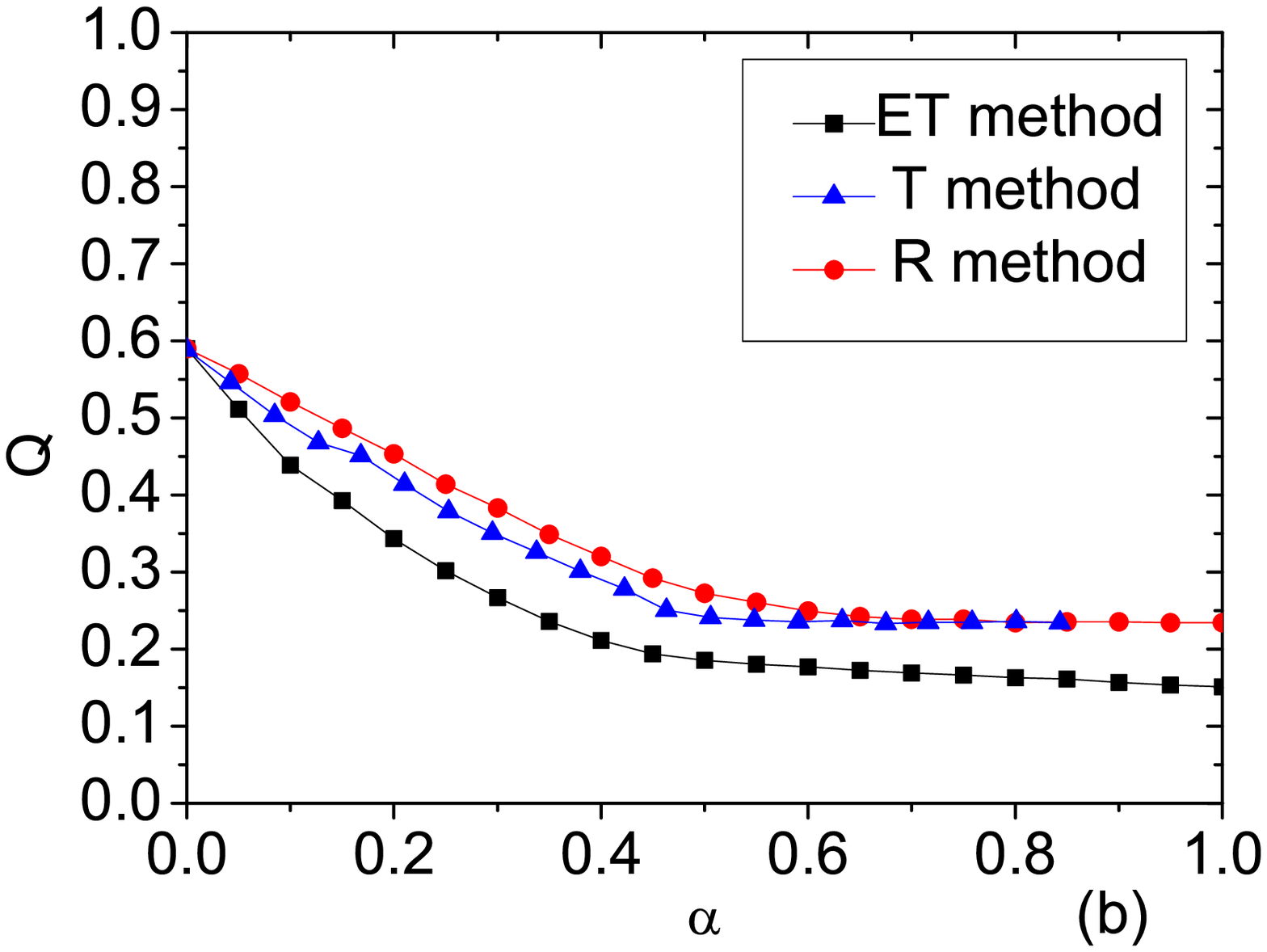}
\includegraphics[width=4cm]{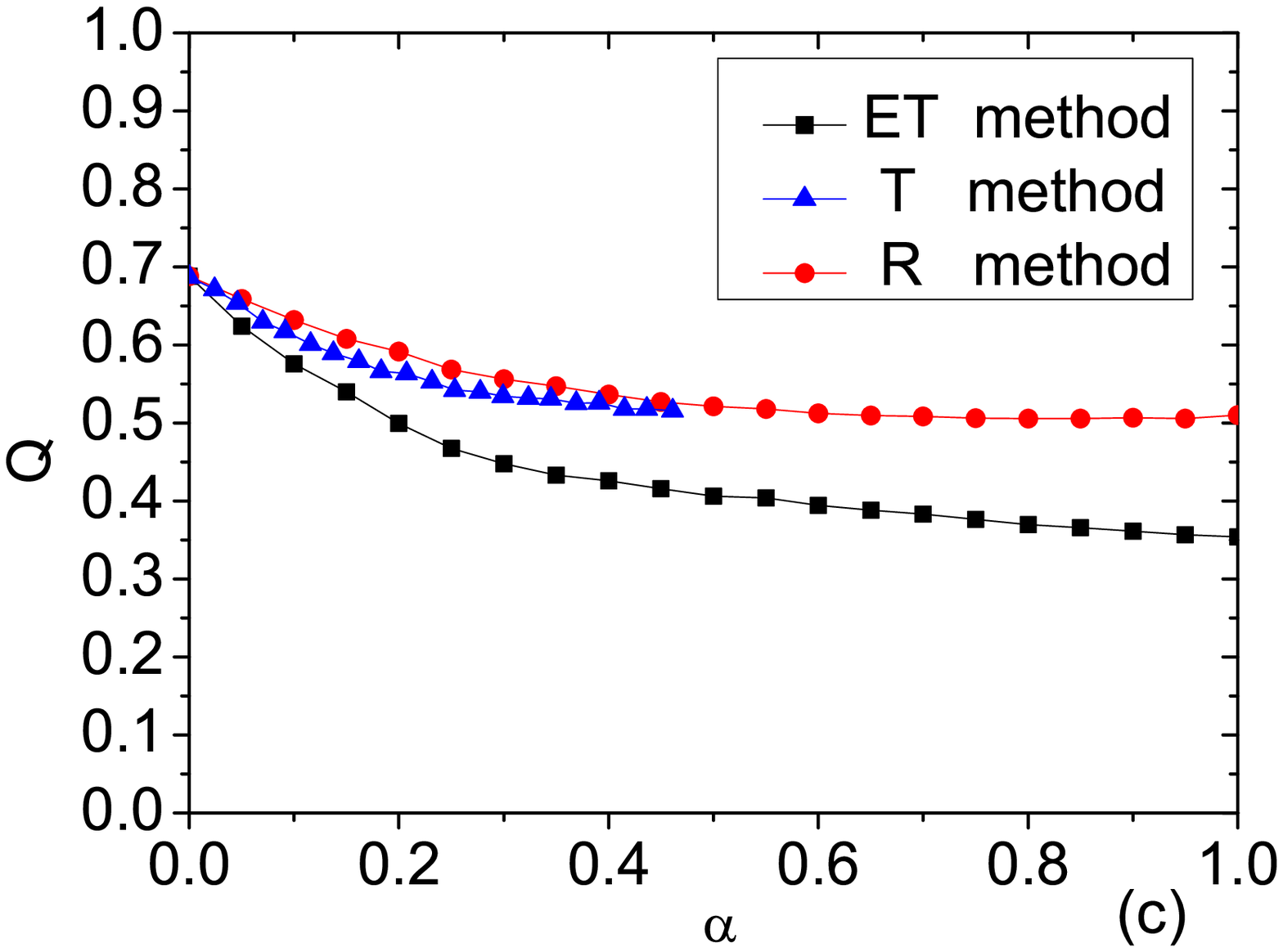}
\caption{ Value of Modularity $Q$ versus $\alpha$. Three attack
methods are simulated on real network, $Q$ decreases with $\alpha$,
which means perturbation makes community structure unclear. (a)
Three attack methods are simulated on Karate club network. (b) Three
attack methods are simulated on football network. (c)Three attack
methods are simulated on econo-physicist network. }\label{fig4}
\end{figure}

When using EO algorithm, modularity analysis is necessary to find
which partition is the best one. The optimal modularity for each
networks, including original one and the disturbed networks, can be
calculated. By doing this, we can measure whether the best community
structure partition for the disturbed network is clear.
Fig.\ref{fig3} and Fig.\ref{fig4} show the modularity changing on
above three artificial and three real networks, caused by the three
different methods introduced above. We can find that modularity
become smaller with the disturbing strengthen going up except the GN
benchmark when $k_{out}$ = 10. The modularity of this artificial
network is low and keeps almost unchanged under targeted attack and
random attack. Meanwhile, the curves of targeted methods depart
significantly from that of random method, indicating that targeted
attacks make the community structure discovered by the algorithm
more obscure than random attack.

\subsection{Edge-clustering coefficient}
\begin{figure}[th]
\centering\includegraphics[width=4cm]{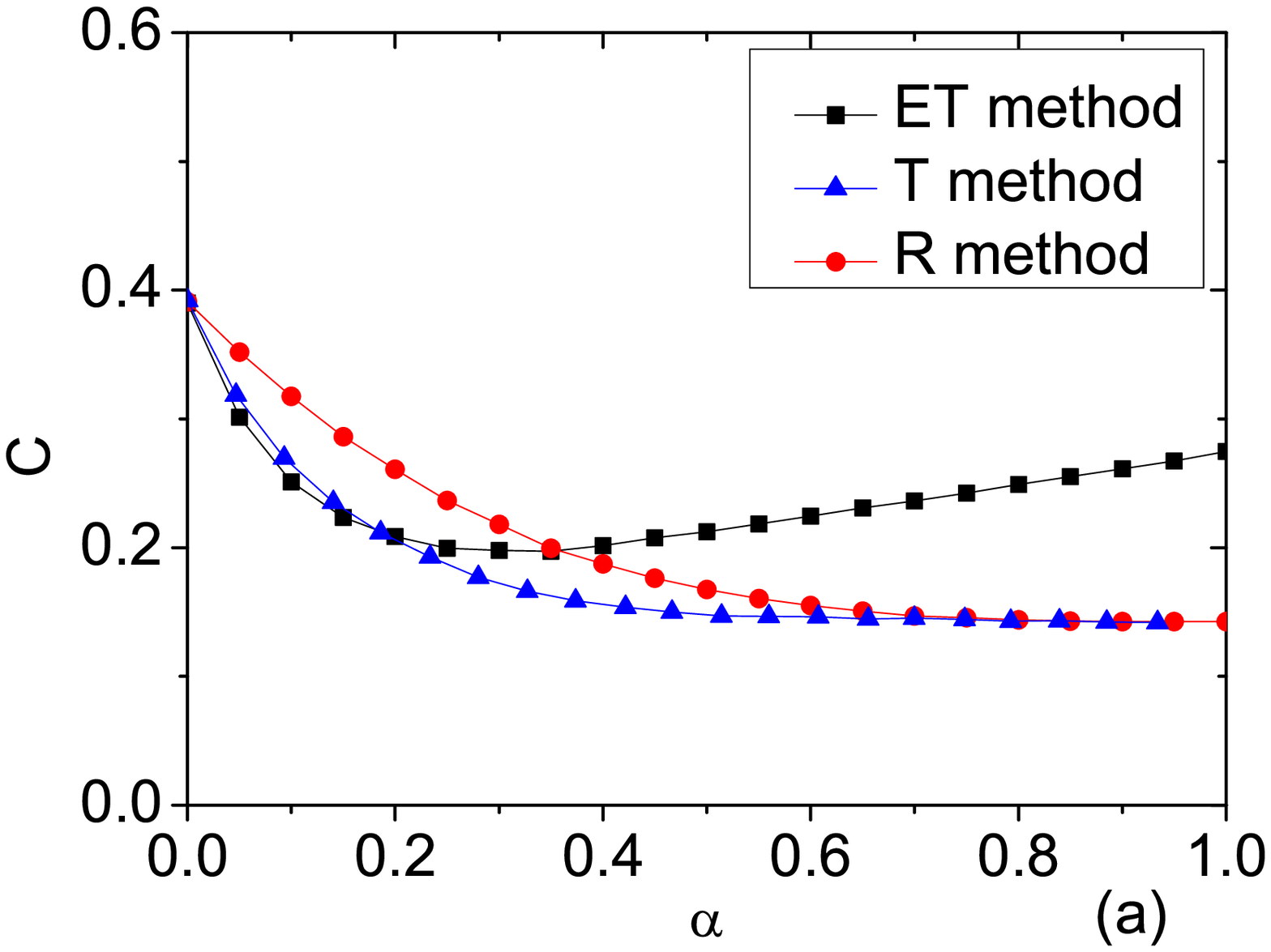}
\includegraphics[width=4cm]{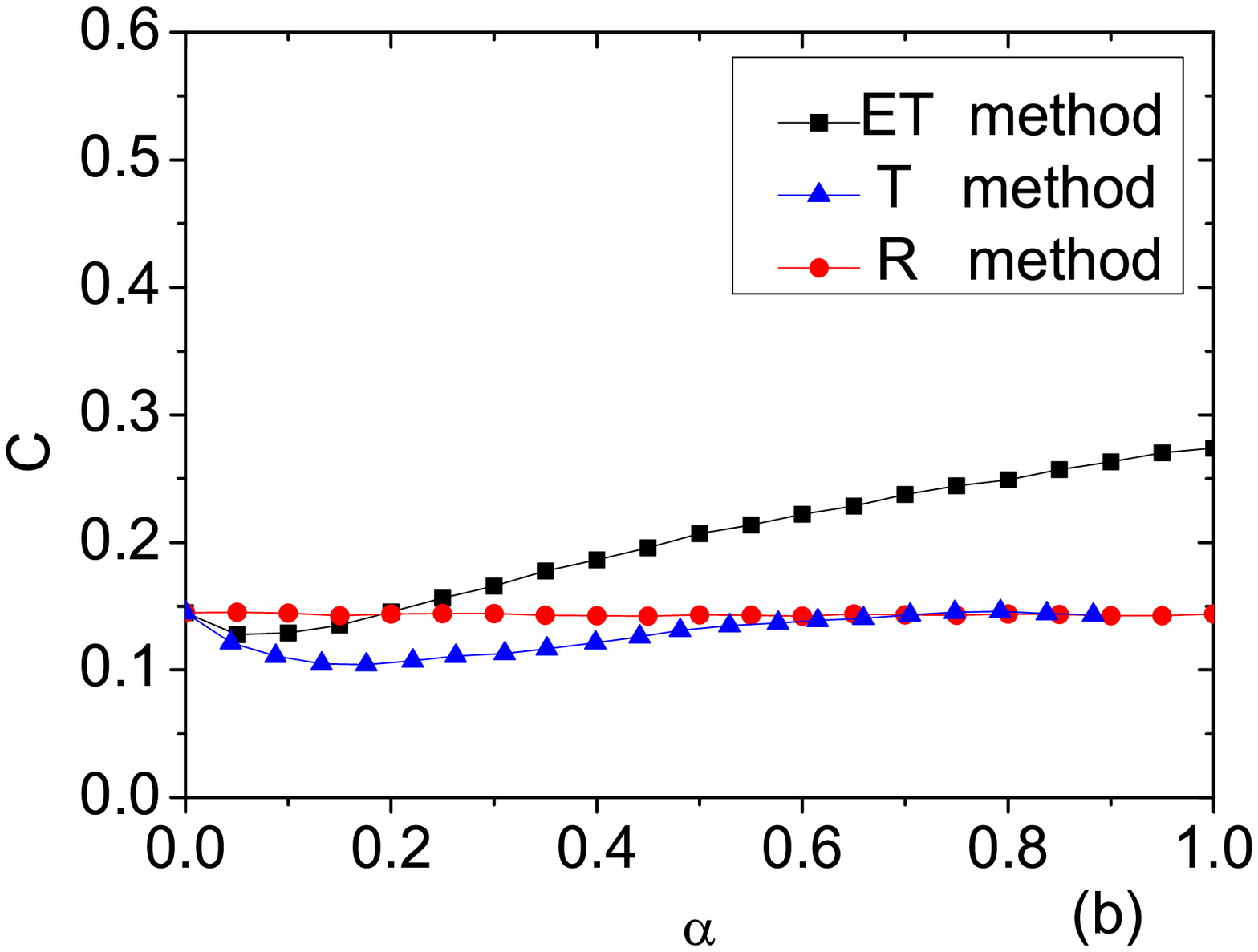}
\includegraphics[width=4cm]{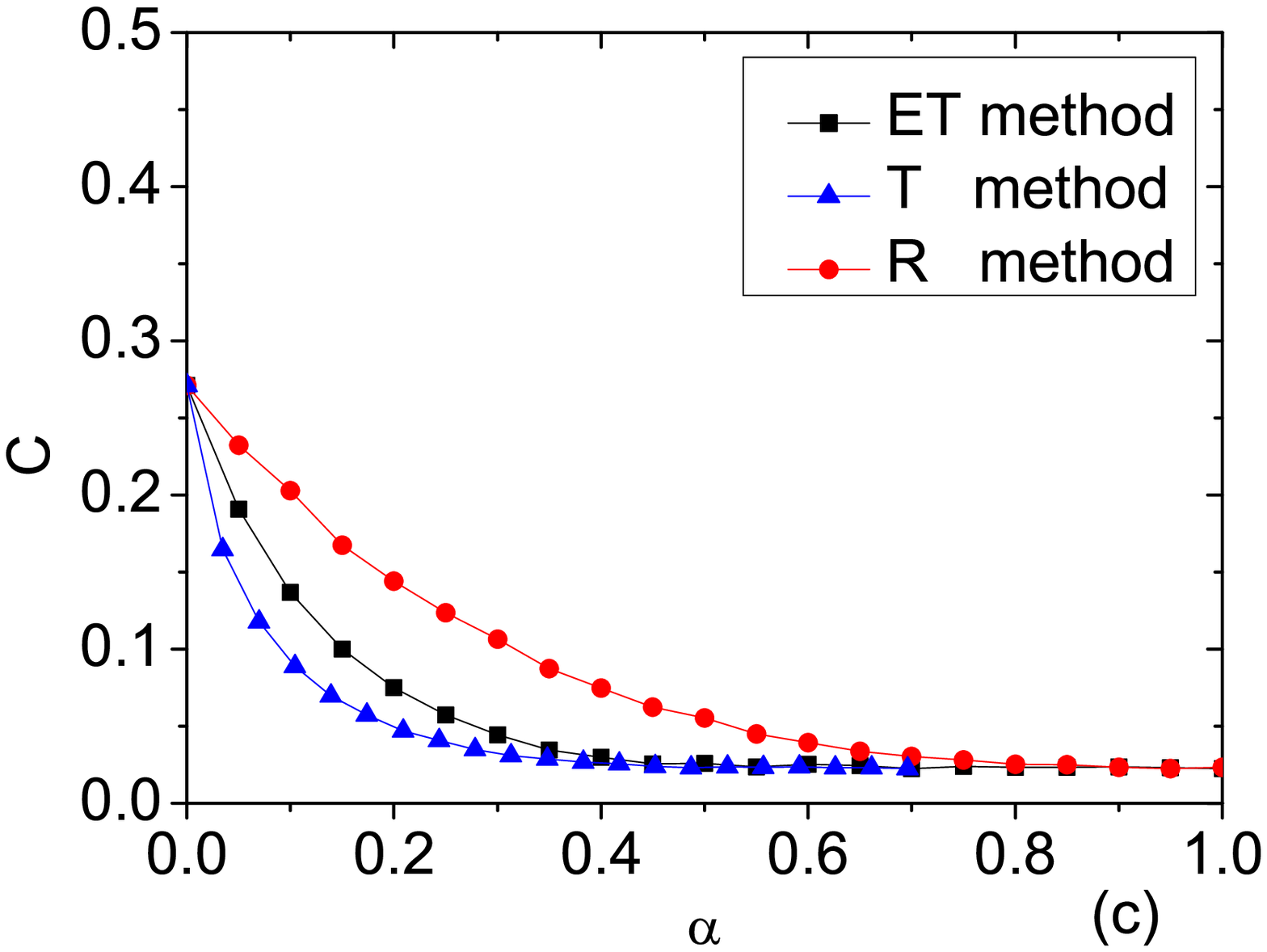}
\caption{Value of edge-clustering coefficient $C$ versus $\alpha$.
Three attack methods are simulated on the arbitrary network
benchmark. (a)$k_{out}$ = 2 on GN benchmark;(b)$k_{out}$ = 10 on GN
benchmark;(c) LFR benchmark, $N$ = 500, $\overline{k}$=10, $p(k)\sim
k^{-\gamma}$, $\gamma = 2.5$, $p(s)\sim k^{-\lambda}$, $\lambda$ =
2.0, mixing parameter $\mu$ = 0.3.}\label{fig5}
\end{figure}

\begin{figure}[th]
\centering\includegraphics[width=4cm]{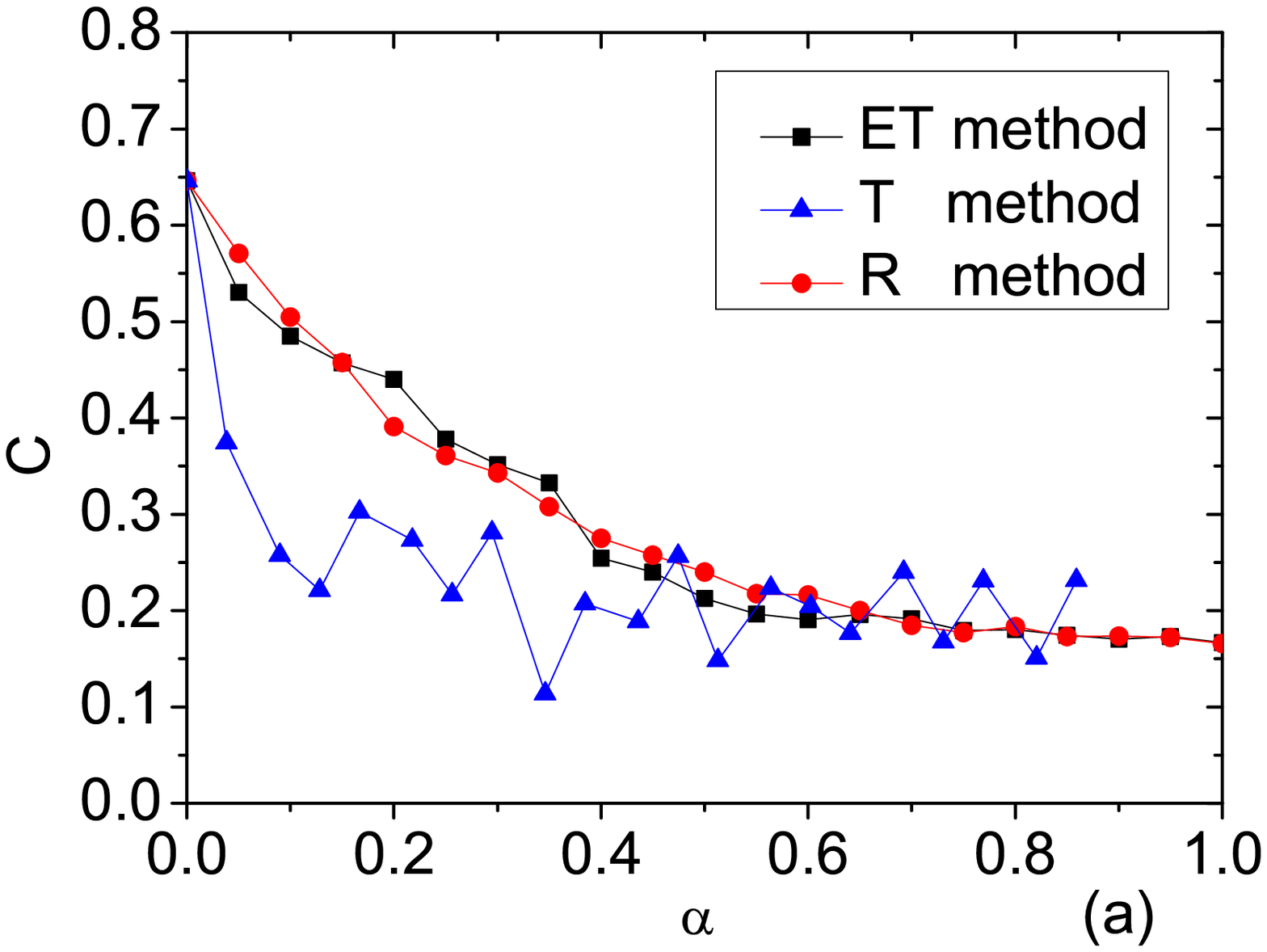}
\includegraphics[width=4cm]{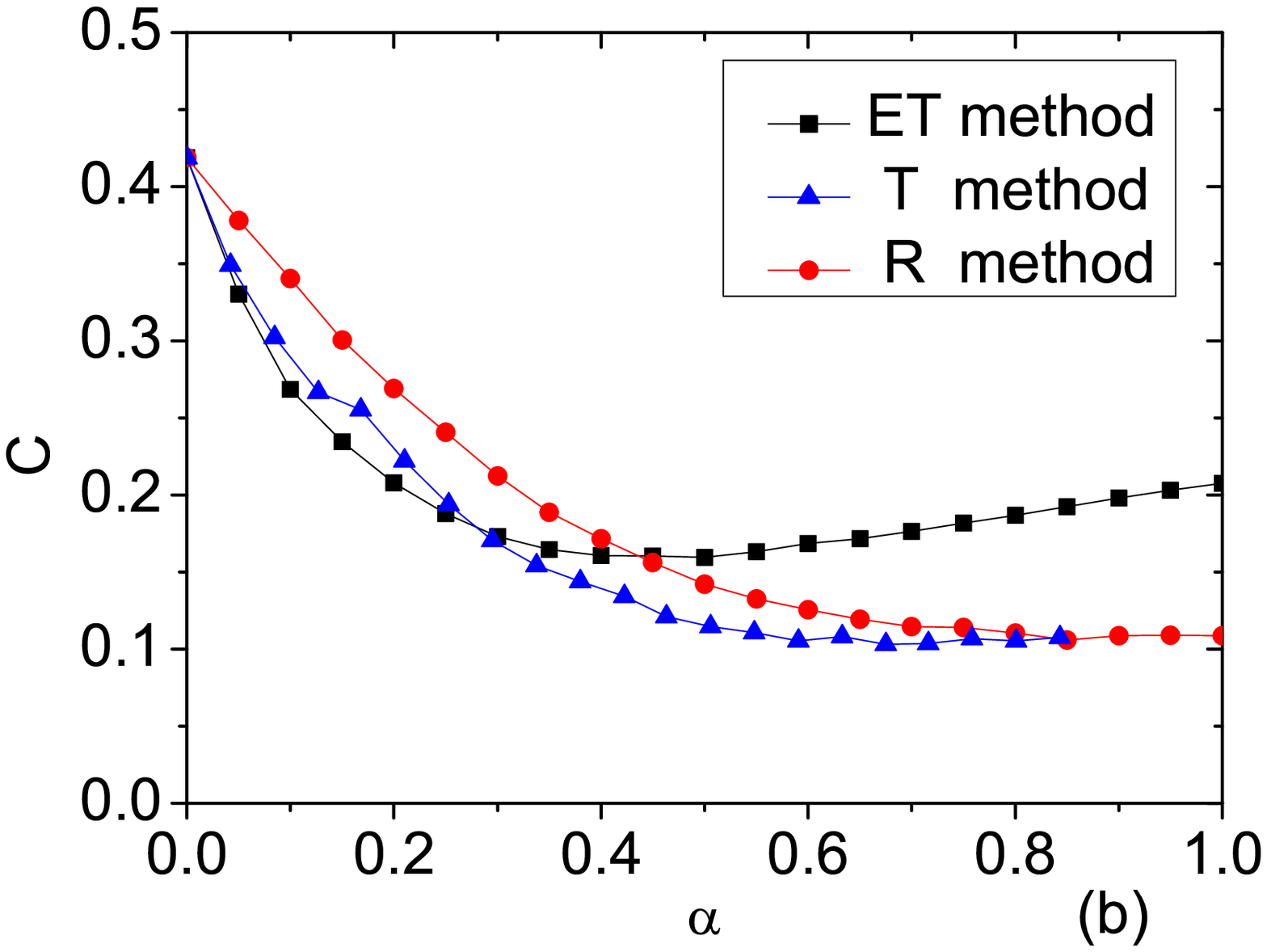}
\includegraphics[width=4cm]{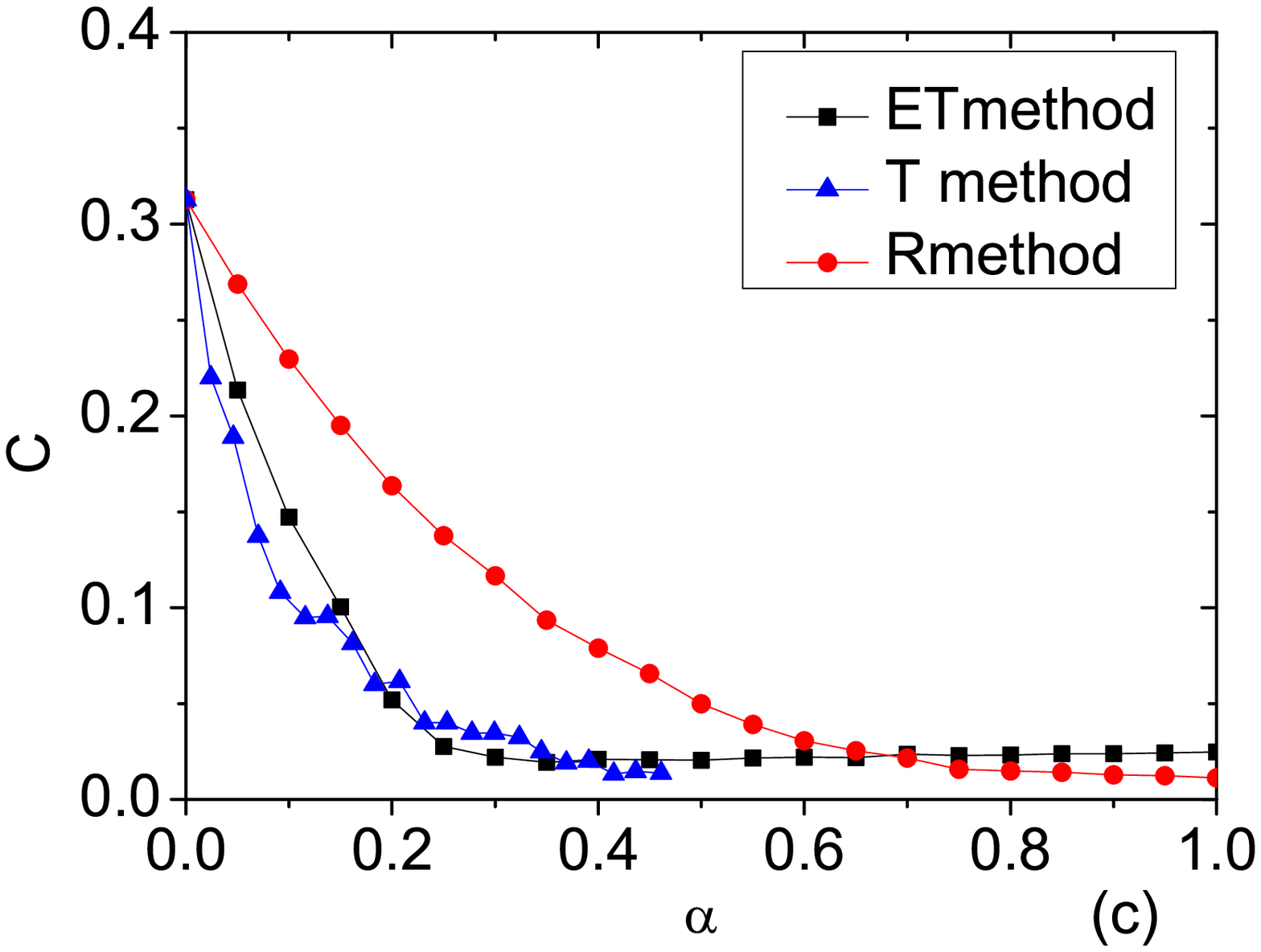}
\caption{Value of edge-clustering coefficient $C$ versus $\alpha$.
Three attack methods are simulated on real network. (a) Three attack
methods are simulated on Karate club network. (b) Three attack
methods are simulated on football network. (c)Three attack methods
are simulated on econophysicist network. }\label{fig6}
\end{figure}

In section \ref{Attack Methods on network}, it has been mentioned
that edge-clustering coefficient is a measurement of the strength of
the network's connection, and network with significant community
structure usually has large edge-clustering coefficient value. In
this part, we mainly investigate the variation of average
edge-clustering coefficient $C$ to perturbation strength $\alpha$.
It is found that in most cases $C$ decreases fast under targeted
methods with $\alpha$'s increasing from zero as shown in
Fig.\ref{fig5} and Fig.\ref{fig6}. Meanwhile, interesting phenomena
appears that on GN benchmark when $k_{out}$ =10 and football
network, that $C$ decreases at the beginning and increases later.

In this part, through modularity and edge-clustering coefficient,
the results suggest that targeted attack method caused more damage
to topology of networks, which is relating to community structure.

\section{conclusion}\label{conclusion}
In the conclusion, we propose targeted methods to disturb the
network, both of which are different from random disturbing. And
then we use the random disturbing method and the our two methods to
disturbing the original network and using function $D$ to compare
the variation of community structure, which is more convenient than
$V$.  The results show that targeted attack methods based on
edge-clustering coefficient and triangles damage more to community
structure through analysis of $D$, $Q$, $C$ than random disturbing.
These facts indicate that community structure is fragile against
targeted attack and imply that triangle is important and deserves
more attention in the study of community structure.

\section*{Acknowledgement}
This work is supported by NSFC under the grant No.$70771011$.

\end{document}